\documentclass[twocolumn,showpacs,preprintnumbers,amsmath,amssymb]{revtex4}

\usepackage{graphicx}
\begin{document}

\title{Current-carrying  molecules: a real space picture}


\author{Anna Painelli}
\email[]{anna.painelli@unipr.it}
\affiliation{Dip. Chimica GIAF, Parma University, \& INSTM UdR Parma,
  43100 Parma, Italy}


\date{\today}

\begin{abstract}
An approach is presented to calculate characteristic current vs
voltage curves for {\it  isolated} molecules without explicit
description of leads. The Hamiltonian for current-carrying
molecules is defined by making resort to Lagrange multipliers,
 while the potential drop needed to sustain the current is calculated
from the dissipated electrical work.  
Continuity
constraints for steady-state DC current  result  in non-linear
potential profiles across the molecule leading, in the adopted
real-space  picture, to a suggestive analogy between 
the molecule  and an electrical circuit.
\end{abstract}


\maketitle


 Experiments on single-molecule
junctions  are challenging, mainly due to the need of contacting a 
 microscopic object, the molecule,  with macroscopic leads 
\cite{reed,cui,reichert,xu}. 
  Theoretical modeling of molecular  junctions is difficult
 \cite{nitzanscience,datta,mukamelpaper,sai}, 
and again the description
of contacts represents a delicate problem.
 To attack the complex problem of conduction through a 
molecular junction a strategy
is emerging \cite{kosov,burke} 
that focuses attention on {\it isolated} molecules and  describes the
intrinsic molecular conductivity in the absence of electrodes.
At variance with common approaches that impose a potential bias 
to the  electrodes and then calculate the resulting  current
 \cite{nitzanscience,datta,mukamelpaper,sai}, 
 a  steady-state DC  current is forced in  the isolated molecule
by making resort to a
 Lagrange-multiplier technique \cite{kosov}, or 
by drawing a  magnetic flux through the molecule
 \cite{burke}.
Whereas the strategy is promising, two main problems remain to be solved:
(1) the calculation  of the potential drop needed to sustain the
current, and (2) the definition of the potential profile in the molecule.
Here I demonstrate that the Joule law can be used to calculate  the
potential drop from the electrical power dissipated on the molecule.
 Moreover, continuity constraints for steady-state DC current 
are implemented in polyatomic
 molecules in terms of {\it multiple}  Lagrange multipliers  that
 yield to   non-linear potential profiles in the molecule.
Finally, in the adopted real-space  picture, the current flows through  
 chemical bonds rather than through energy levels, 
leading to a suggestive description of the 
molecule as  an electrical circuit with resistances
associated to chemical bonds.

To start with consider a  diatomic Hubbard molecule, whose Hamiltonian
$H_0$ is defined by $U$, $t$, and the difference of on-site energies: 
$2\Delta=\epsilon_2-\epsilon_1$.
Following Kosov \cite{kosov},   a current is forced through the molecule
by introducing a  Lagrange multiplier, $\lambda$, as follows:
\begin{equation}
  H(\lambda)= H_0 -\lambda \hat j
\label{hj}
\end{equation}
where $c^\dagger_{i,\sigma}$ creates an electron with 
spin $\sigma$ on the $i$-site, and 
 $\hat j=-it\sum_\sigma(c^\dagger_{1\sigma} c_{2\sigma} -H.c.)$ 
measures the current flowing through  the bond. 
Here and in the following  $\hbar$ and the electronic
 charge are set
to 1, and  $t$ is taken as the  energy unit.
The ground state of $H(\lambda)$, $|G(\lambda)\rangle$, carries
a finite current, $J=\langle G(\lambda)| \hat j|G(\lambda)\rangle$, and
the Lagrange multiplier, $\lambda$, is fixed by imposing a predefined
$J$  \cite{kosov}. 
Other  molecular properties can be calculated  as well, 
 and  their dependence on $J$ can be investigated
\cite{kosov}. Just
as an example, the bond-order decreases with $J$, and,
in systems with inequivalent sites, the on-site charge
distribution is equalized by the current flow.
These are interesting informations, but characteristic $J(V)$ curves
are still needed.

The Lagrange multiplier, $\lambda$,  has the
 dimensions and the meaning of a magnetic flux drawn across
the molecule  to generate a spatially uniform electric
field \cite{kohn,burke},  
$E \propto \omega\lambda$, where $\omega $ is the field frequency 
\cite{kohn}.
In the limit of  static fields,  $\omega \rightarrow 0$, both $E$ and
 $V$   vanish, suggesting that a
finite current flows  in the molecule at zero bias. 
This contrasts sharply  with the  fundamental
relation  between charge transport and energy dissipation
\cite{nitzanreview,datta,burke}: a finite $V$ is needed to sustain a current
due to dissipative phenomena occurring in the conductor. 
Specifically, the Joule law  relates
 the potential drop in a conductor to the electrical power spent on
 the system to sustain the current,  $W=VJ$. Since $J$ is known, $V$ 
 can be obtained from a calculation of the dissipated power.

Dissipation is conveniently described in the density matrix
formalism using as a  basis of the eigenstates  $|k\rangle$ of  $H(0)$
\cite{mukamel,boyd}. 
The equilibrium  density matrix, $\sigma_0$, is a diagonal matrix
 whose elements are fixed by  the  Boltzmann distribution.
 On the same basis  $\sigma(\lambda)$ is a non-diagonal
matrix corresponding to a  non-equilibrium state whose dynamics is
governed by: 
$\dot \sigma= -\frac{i}{\hbar}[H,\sigma] +\dot \sigma_R$, where
$\dot \sigma_R$ accounts for relaxation phenomena, as  due to 
 all degrees of freedom not explicitly described by
$H$ (e.g. molecular vibrations, or environmental degrees of freedom
also including leads) \cite{mukamel,datta,nitzanreview}.
Diagonal elements of $\dot \sigma_R$ 
describe depopulation and are associated with energy dissipation.
As for depopulation  I adopt a simple phenomenological model with 
$(\dot \sigma_R)_{kk} =  -\sum_m\gamma_{km} \sigma_{mm}$, where
$\gamma_{km}$ measures the probability of the transition from $k$ to
$m$ \cite{mukamel,boyd}.
For the sake of simplicity I will consider  
the  low-temperature limit, with 
$\gamma_{km}=\gamma$ for $k>m$ and
$\gamma_{km}=0$ otherwise. 
The dynamics of off-diagonal elements is governed by
depopulation and dephasing effects: $(\dot \sigma_R)_{km}=
-\Gamma_{km}\sigma_{km}$, with $\Gamma_{km}=
(\gamma_{kk}+\gamma_{mm})/2 +\gamma'_{km}$, where
$\gamma_{kk}=\sum'_m\gamma_{mk}$, and $\gamma'_{km}$ describes 
dephasing, i.e. the loss of coherence due to elastic
scattering \cite{mukamel,boyd}.

The energy dissipated by the system, $Tr{(\dot \sigma_r H)}$,
has two contributions: the first one,
$W_d=\sum_k(\dot\sigma_R)_{kk}$,  is always negative and measures the
energy that the system dissipates to the bath as the current
flows. This term is governed by depopulation, whereas
dephasing plays no role. The second term, $W=-\lambda Tr(
\dot\sigma_R\hat j)$,
measures the electric work done on the system to sustain the current: 
it is this term that
enters the Joule law. In non-degenerate systems  $\hat j$ is an 
off-diagonal operator, so that only off-diagonal elements of $\dot \sigma_{R}$
enter the expression for $W$. Both depopulation and dephasing
then  contribute to $V$, and hence  to the
molecular resistance. This is in line with the  observation
that a current flowing through a molecule implies an  organized motion of
electrons along a specific direction \cite{buttiker,nitzanreview,datta}.
 Therefore any mechanism of scattering, either anelastic, as described
 by depopulation, or elastic, as described by dephasing, contributes to
 the electrical resistance \cite{buttiker,datta}. 
The unbalance between $W_d$ and $W$ is always
positive:  the molecule heats as current flows. Efficient  heat
dissipation is fundamental to reach a steady-state regime and to avoid
molecular  decomposition \cite{nitzanreview}.

Fig. 1 shows the characteristic curves calculated for a diatomic
 molecule  with $\gamma =0.2$ and $\gamma'_{km}=0$.
In the left panel results are shown
for the symmetric, $\Delta=0$, system. As expected, electronic
correlations decrease the conductivity. The results
in the right panel for an asymmetric system ($\Delta \ne 0$) show
instead an
increase of the low-voltage 
conductivity with increasing $U$. This  interesting result is
related to the minimum excitation gap, and hence the maximum
 conductance, of the  system with $U=2\Delta$.
\begin{figure}
\includegraphics* [scale=0.45]{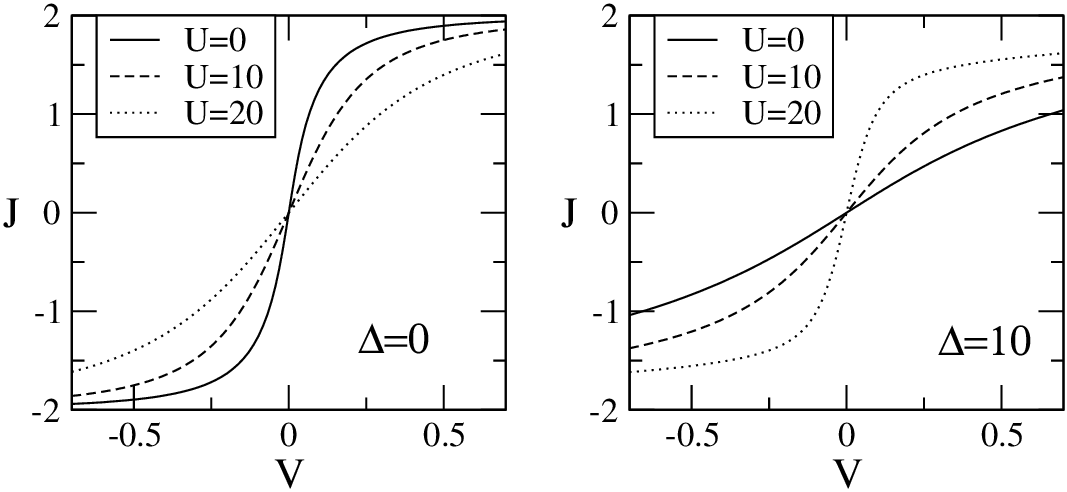}
\caption{Characteristic $J(V)$ curves for a two-site system with  
different $\Delta$ and $U$. The depopulation rate $\gamma$ is set to
0.2, dephasing is neglected.\label{figura1}}
\end{figure}
The asymmetric diatomic molecule  represents  a minimal model for 
the Aviram and Ratner rectifier \cite{aviramratner}, however 
the characteristic curves in the right panel of Fig. \ref{figura1}
 are symmetric, and do not support rectification. In
 agreement with recent results,
 rectification in asymmetric molecules 
is most probably due to contacts \cite{datta}, or  to the
coupling between  electrons and 
 vibrational or conformational degrees degrees of freedom \cite{troisi}.

Before attacking the more complex problem of polyatomic molecules, it
is important  to compare the results obtained so far 
 with well known results for the optical  conductivity \cite{kohn}. 
If   $\Gamma_{km}=\Gamma$, as it
occurs, e.g., for systems with large  inhomogeneous broadening 
 (the coherent conductance limit \cite{nitzanscience,nitzanreview}),
 the expression for  the potential drop is 
very simple: $V=\lambda \Gamma$.  Then, a perturbative
expansion of $J$ leads to the following expression for the zero-bias 
conductivity, $\mathcal{G}_0$:
\begin{equation}
 {\mathcal  G}_0=\frac{2}{\Gamma} \sum_k\frac{\langle k |\hat j |g\rangle|^2}{E_k-E_g}
\end{equation}
where $g$ is the gs of $H(0)$, with energy $E_g$, 
 and the sum runs on all excited states.
This expression for the DC conductivity coincides with the 
zero-frequency limit of the  optical conductivity \cite{kohn},
provided that the frequency, $\omega$, appearing in the
denominator of the expression for the optical conductivity in
 Ref. \cite{kohn} is substituted  by 
$\omega-i\Gamma$. 
 Introducing  a complex frequency to account for relaxation
is a standard procedure in spectroscopy \cite{boyd}, leading to
similar effects as the 
introduction of an exponential  switching on of the electromagnetic
field  \cite{kohn}: both phenomena account for the loss of coherence
of  electrons driven by an EM field and properly suppress
the divergence  of the optical conductivity due to
the build-up of the phase of electrons driven by a static field.

The connection between  DC and optical conductivity breaks down in
polyatomic molecules. To keep the discussion simple, I will focus
attention on  linear Hubbard chains. The  optical conductivity 
of Hubbard chains was
discussed based on the  current operator
 $\hat J=-ie\sum_it_i(c^\dagger_{i,\sigma} c_{i+1,\sigma} -H.c.)/\hbar$
\cite{maldague}. However, this operator 
 measures the {\it average} total current and 
 does not apply to DC currents.
Specifically, if
 a term $-\lambda \hat J$ is added to the molecular Hamiltonian, 
a finite average current is forced through the molecule
\cite{kosov}, but this current does not satisfy 
basic continuity constraints for steady-state DC current.
In fact, to sustain a steady-state DC current one must avoid the
build  up of electrical 
 charge at atomic sites. 
Specifically, in linear molecules  the continuity constraint imposes
that exactly the same amount of current
flows through  each bond in the molecule. To impose this
constraint  the current on each single bond must be under control and
   a Lagrange  multiplier must be introduced for  each
bond, as follows:
\begin{equation}
  H(\lambda_i)= H_0 -\sum_i \lambda_i \hat j_i
\label{hji}
\end{equation}
where $\hat j_i=-it_i(c^\dagger_{i,\sigma} c_{i+1,\sigma}
-H.c.)/\hbar$, 
and the $\lambda_i$'s are fixed by imposing $j_i=\langle
G|\hat j_i|G\rangle=J$  independent on $i$.

As before,  the electrical
work done on the molecule is:
\begin{equation}
  W=-\sum_i \lambda_i Tr(\hat j_i \dot \sigma_R)
\end{equation}
that  naturally separates into contributions, $W_i$, relevant to each
bond. The total potential drop across the molecule, $V=W/J$, is then
the sum of the potential  drops across each bond, 
$V_i= W_i/J$, leading in general to non-linear potential
profiles.  Of course, in the adopted real-space
picture the potential profile can only be calculated at atomic
positions, and 
no information can be obtained on the potential profile inside each
bond. Therefore,  instead of showing the potential profile along the
molecule, I prefer to convey the same information in terms 
 of bond-resistances, defined as: $R_i=(\partial J/\partial V_i)^{-1}$.
\begin{figure}
\includegraphics* [scale=0.49]{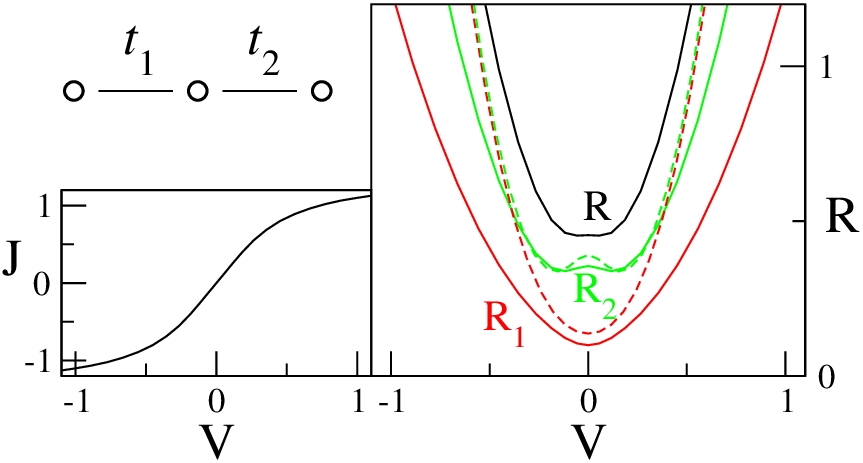}
\caption{\label{3homo} Left panel: characteristic curve for the
  three-site Hubbard chain sketched in the figure,
 with 3 electrons, $U=4$, constant on-site
  energies and $t_1=1.2$, $t_2=0.8$. Dephasing is set to zero, and
  depopulation rate is 
$\gamma=0.1$. Right panel:  total resistance ($R$)  and
  bond-resistances
($R_1$ and $R_2$). For
  bond-resistances continuous and dashed lines show results obtained
  by allowing the current to flow through the whole molecule, and
  through a single bond, respectively.}
\end{figure}
Fig. \ref{3homo} shows the behavior
of a 3-site chain with three electrons, $U=4$, equal on-site energies
and different $t$: $t_1=1.2$, and $t_2=0.8$. The left panel shows the
characteristic $J(V)$ curve, and continuous lines in the right  panel
report the total resistance $R$, and the two bond resistances,  $R_1$
and  $R_2$. The molecular  resistance varies with the applied voltage
and, as expected, the resistance of the  weaker
bond is higher than the resistance of the stronger bond.
 
Dimensionless resistances in the figure are in units with
$\hbar/e^2=(2\pi\tilde g_0)^{-1}=1$, where $\tilde g_0$ is the quantum of
conductance, that, in standard approaches  to molecular junctions \cite{datta} 
represents the maximum conductance associated with  a discrete
molecular  level. This well known result is
 related to the inhomogeneous broadening of molecular
energy levels as due to
their  interaction with the electrodes \cite{datta}. Of course there
is no {\it intrinsic} limit to the conductivity in the model for
 isolated molecules discussed here.

As a direct consequence of the continuity constraint, 
the total resistance $R=(\partial J /\partial V)^{-1}$ is the sum of
the two bond-resistances, leading to a suggestive 
description of the  molecule as an
electrical circuit, with resistances associated with chemical bonds
 joint in series at
atomic sites. Whereas this picture is useful,  the concept
 of  bond-resistance should be considered with care in
molecular circuits. At
variance with standard conductors, in fact, the resistance of the bonds depends
not only on the circuit (the molecule) they are inserted in, but also
on the way the resistance is measured. Dashed lines in
Fig. \ref{3homo} show the bond-resistances
calculated by  forcing the current through specific bonds 
(i.e. by setting a single $\lambda_i\ne 0
$ in Eq. \ref{hji}), 
 and these differ from  the bond-resistances calculated
when the whole molecule carries the current (continuous lines).

\begin{figure}
\includegraphics* [scale=0.49]{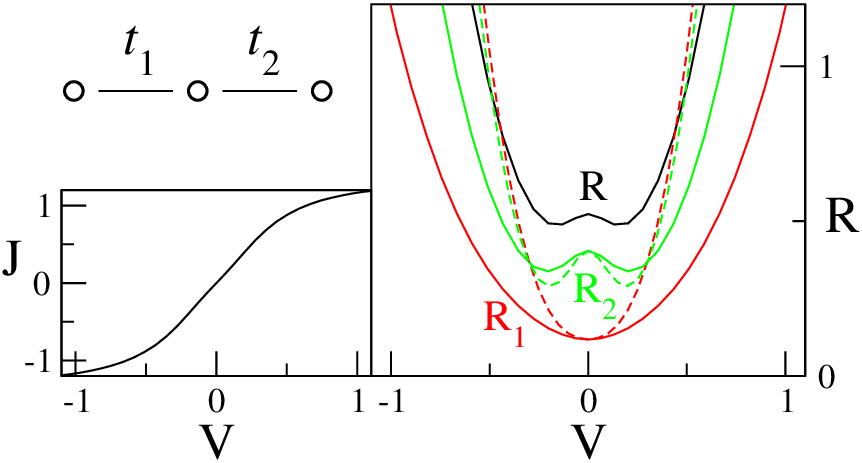}
\caption{\label{3in} The same as in fig.\ref{3homo}, but with 
  $\gamma=0$ and   $\gamma'_{km}=0.2$.}
\end{figure}

The situation becomes somewhat simpler in the coherent conductance
limit, $\Gamma_{km}=\Gamma$. Perturbative arguments 
can be used to demonstrate that at zero bias the bond resistances
calculated for  the current flowing through the whole molecule or
through a single bond do coincide. 
This {\it additive} result for the molecular resistance in the coherent 
transport limit is  in line with the observation of transmission rates 
inversely proportional to the molecular length in the same limit \cite{davis}. 
However,  as shown in 
Fig. \ref{3in}, this simple {\it Ohmic} behavior  breaks down quickly at finite
bias.

The introduction of as many Lagrange multipliers as many 
current-channels (bonds) are present in the molecule accounts for
a non-linear potential profile through the molecule, i.e. for a
non-uniform electric field. Accounting for  a single Lagrange multiplier 
coupled to the total current operator is equivalent to 
draw a magnetic flux through the molecule as to generate   a {\it spatially
  homogeneous} electric field
\cite{kohn,maldague,burke}, a poor approximation for DC conductivity 
 in extended (polyatomic) molecules.
 Just as an example, for a  4-site chain with
the same $t_i=1$ on each bond,  the 
zero-bias resistance of the
central bond exceeds that of the lateral bonds 
with $R_2/R_1$ ranging from   10 to 2 as $U$
increases from 0 to 4. Bonds with the same $t$ have different
resistances due to their different bond-orders, and,
 in agreement with recent results
\cite{dattaprofiles,mukamelpaper}, this  demonstrates nicely  the need of
accounting  for non-uniform electric fields in extended molecules, even for
very idealized molecular structures.

It is of course  possible to discuss more complex molecular models. As
an interesting example,  a nearest-neighbor
hopping $t'$ is added to the Hamiltonian  for the three site molecule discussed above.
 This opens a new channel for electrical transport, and   a
term $-\lambda'\hat j'$ adds to the Hamiltonian with
 $\hat j'=-it'\sum_\sigma(c^\dagger_{1\sigma}
c_{3\sigma} -H.c.)$.
 As before, continuity imposes $j_1=j_2$, as to avoid building up  of
charge at the central site. The total current is $J=j_1+j'$ and, 
 of course, no continuity constraint is given on  $j'$.
 However the potential drop across the molecule, i.e. the
potential drop measured at sites 1 and 3 must be uniquely defined.
 Therefore one
must tune $\lambda_1$, $\lambda_2$ and $\lambda'$ as to satisfy
$j_1=j_2$, while satisfying the condition: $V_1+V_2=V'=V$, with
$V_{i}=W_{i}/j_{i}$ and $V'=W'/j'$. Imposing a constraint on the potentials
 is a tricky affair, that becomes trivial when
$\Gamma_{km}=\Gamma$. 
\begin{figure}
\includegraphics* [scale=0.49]{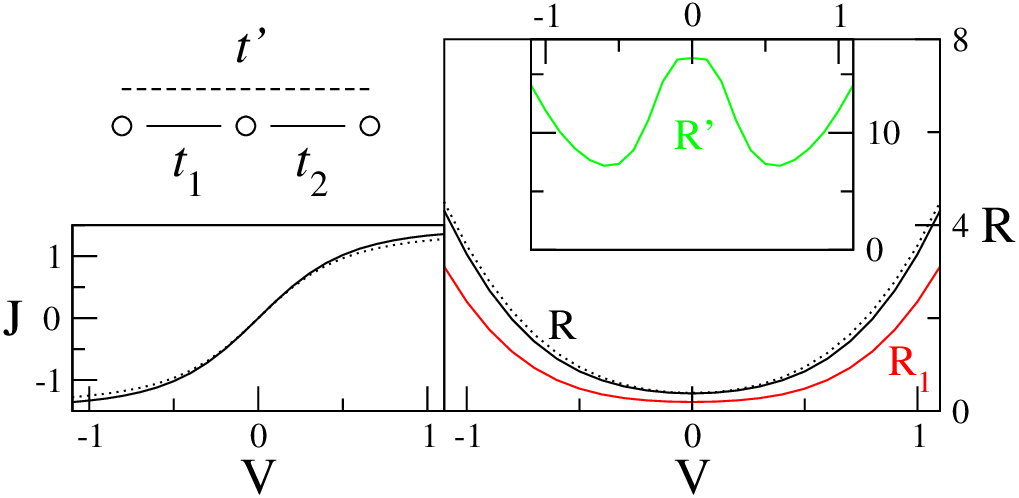}
\caption{Left panel: characteristic curves  of 
the three site  chain sketched in the figure, 
with three electrons, constant on-site energies,
 $U=4$, $t_1=t_2=1$
 and $\gamma_{km}=0$ and $\gamma'_{km}=0.2$. Continuous and dotted
 lines refer to  $t'=0.4$
 and 0, respectively. Right panel: 
 molecular and bond-resistances 
 for the chain with $t'=0.4$. The dotted line shows the total
 resistance for the chain with $t'=0$. 
\label{ponte}}
\end{figure}
In that case in fact $V_i=\Gamma \lambda_i$ and $V'=\Gamma \lambda'$,
so that  the
constraint on the potentials  immediately translates into
a constraint on Lagrange multipliers. Fig. \ref{ponte} shows some results
obtained in this limit for a system with $t_1=t_2=1$, $t'=0.4$. 
In spite of the fairly large $t'$ value,  the contribution to the
current from the bridge-channel is small, mainly due to the small
bond-order for next-nearest neighbor sites. 
Once again the physical constraints imposed to the currents and to the
potentials lead to standard combination rules for
bond-resistances with $1/R=1/R' +1/(R_1+R_2)$. As for the DC
conductivity is concerned, the molecule behaves as an electrical
circuit with two resistances, $R_1$ and $R_2$  in series bridged by a parallel
resistance, $R'$.

Applying the proposed 
 approach to complex molecular structures  and/or to 
 molecules described by accurate quantum chemical  Hamiltonians  
 is non-trivial
due to the appearance in the Hamiltonian of as many
 Lagrange multipliers  as many current channels are
considered, and due  to the large number of constraints to
be implemented. Instead,  at least for small molecules,  the approach 
can be fairly easily extended to account for  vibrational degrees of
freedom. Non-adiabatic calculations are currently in progress to describe
 electrical conduction through a diatomic 
molecule in the presence of Holstein and Peierls electron-phonon
 coupling. More refined models for the
 relaxation dynamics can also be implemented \cite{davis},
 whereas the introduction of
 spin-orbit coupling can lead to a model for  spintronics.

In conclusion, this paper presents an  approach to the calculation of
characteristic current/voltage curves for isolated molecules in the
absence of contacts. While hindering the direct comparison with
experimental data, this allows the definition of the molecular
conductivity as an intrinsic molecular property. Even more important, a
paradigm is defined for imposing a steady-state DC current through the
molecule, while extracting the voltage drop across the molecule from
the energy dissipation. 
The careful implementation of continuity
constraints for steady-state DC current leads to the definition
of the potential profile through the molecule, that, in the adopted
real-space description, quite naturally results in  the
concept of bond-resistances, in a suggestive description of the
molecule as an electrical circuit with current flowing through
chemical bonds.

I thank D. Kosov for   useful discussions and
correspondence. Discussions with A. Girlando, S. Pati, S. Ramasesha,
and Z.G. Soos are gratefully acknowledged. 
The contributions from
S. Cavalca and C. Sissa in the early stages of this work are
acknowledged. 
Work supported by Italian  MIUR through FIRB-RBNE01P4JF
and PRIN2004033197-002. 


\bibliography{me}

\begin{thebibliography}{20}
\expandafter\ifx\csname natexlab\endcsname\relax\def\natexlab#1{#1}\fi
\expandafter\ifx\csname bibnamefont\endcsname\relax
  \def\bibnamefont#1{#1}\fi
\expandafter\ifx\csname bibfnamefont\endcsname\relax
  \def\bibfnamefont#1{#1}\fi
\expandafter\ifx\csname citenamefont\endcsname\relax
  \def\citenamefont#1{#1}\fi
\expandafter\ifx\csname url\endcsname\relax
  \def\url#1{\texttt{#1}}\fi
\expandafter\ifx\csname urlprefix\endcsname\relax\def\urlprefix{URL }\fi
\providecommand{\bibinfo}[2]{#2}
\providecommand{\eprint}[2][]{\url{#2}}

\bibitem[{\citenamefont{Reed et~al.}(1997)\citenamefont{Reed, Zhou, Muller,
  Burgin, and Tour}}]{reed}
\bibinfo{author}{\bibfnamefont{M.~A.} \bibnamefont{Reed}},
  \bibinfo{author}{\bibfnamefont{C.}~\bibnamefont{Zhou}},
  \bibinfo{author}{\bibfnamefont{C.~J.} \bibnamefont{Muller}},
  \bibinfo{author}{\bibfnamefont{T.~P.} \bibnamefont{Burgin}},
  \bibnamefont{and} \bibinfo{author}{\bibfnamefont{J.~M.} \bibnamefont{Tour}},
  \bibinfo{journal}{Science} \textbf{\bibinfo{volume}{278}},
  \bibinfo{pages}{252} (\bibinfo{year}{1997}).

\bibitem[{\citenamefont{Cui et~al.}(2001)\citenamefont{Cui, Primak, Zarate,
  Tomfohr, Sankey, Moore, Moore, Gust, Harris, and Lindsay}}]{cui}
\bibinfo{author}{\bibfnamefont{X.~D.} \bibnamefont{Cui}},
  \bibinfo{author}{\bibfnamefont{A.}~\bibnamefont{Primak}},
  \bibinfo{author}{\bibfnamefont{X.}~\bibnamefont{Zarate}},
  \bibinfo{author}{\bibfnamefont{J.}~\bibnamefont{Tomfohr}},
  \bibinfo{author}{\bibfnamefont{O.~F.} \bibnamefont{Sankey}},
  \bibinfo{author}{\bibfnamefont{A.~L.} \bibnamefont{Moore}},
  \bibinfo{author}{\bibfnamefont{T.~A.} \bibnamefont{Moore}},
  \bibinfo{author}{\bibfnamefont{D.}~\bibnamefont{Gust}},
  \bibinfo{author}{\bibfnamefont{G.}~\bibnamefont{Harris}}, \bibnamefont{and}
  \bibinfo{author}{\bibfnamefont{S.~M.} \bibnamefont{Lindsay}},
  \bibinfo{journal}{Science} \textbf{\bibinfo{volume}{294}},
  \bibinfo{pages}{571} (\bibinfo{year}{2001}).

\bibitem[{\citenamefont{Reichert et~al.}(2002)\citenamefont{Reichert, Ochs,
  Beckmann, Weber, Mayor, and Lohneysen}}]{reichert}
\bibinfo{author}{\bibfnamefont{J.}~\bibnamefont{Reichert}},
  \bibinfo{author}{\bibfnamefont{R.}~\bibnamefont{Ochs}},
  \bibinfo{author}{\bibfnamefont{D.}~\bibnamefont{Beckmann}},
  \bibinfo{author}{\bibfnamefont{H.~B.} \bibnamefont{Weber}},
  \bibinfo{author}{\bibfnamefont{M.}~\bibnamefont{Mayor}}, \bibnamefont{and}
  \bibinfo{author}{\bibfnamefont{H.~V.} \bibnamefont{Lohneysen}},
  \bibinfo{journal}{Phys. Rev. Lett.} \textbf{\bibinfo{volume}{88}},
  \bibinfo{pages}{176804} (\bibinfo{year}{2002}).

\bibitem[{\citenamefont{Xu and Tao}(2003)}]{xu}
\bibinfo{author}{\bibfnamefont{B.~Q.} \bibnamefont{Xu}} \bibnamefont{and}
  \bibinfo{author}{\bibfnamefont{N.~J.} \bibnamefont{Tao}},
  \bibinfo{journal}{Science} \textbf{\bibinfo{volume}{301}},
  \bibinfo{pages}{1221} (\bibinfo{year}{2003}).

\bibitem[{\citenamefont{Nitzan and Ratner}(2003)}]{nitzanscience}
\bibinfo{author}{\bibfnamefont{A.}~\bibnamefont{Nitzan}} \bibnamefont{and}
  \bibinfo{author}{\bibfnamefont{M.~A.} \bibnamefont{Ratner}},
  \bibinfo{journal}{Science} \textbf{\bibinfo{volume}{300}},
  \bibinfo{pages}{1384} (\bibinfo{year}{2003}).

\bibitem[{\citenamefont{Datta}(2004)}]{datta}
\bibinfo{author}{\bibfnamefont{S.}~\bibnamefont{Datta}},
  \bibinfo{journal}{Nanotechnology} \textbf{\bibinfo{volume}{15}},
  \bibinfo{pages}{S433} (\bibinfo{year}{2004}).

\bibitem[{\citenamefont{Berman and Mukamel}(2004)}]{mukamelpaper}
\bibinfo{author}{\bibfnamefont{O.}~\bibnamefont{Berman}} \bibnamefont{and}
  \bibinfo{author}{\bibfnamefont{S.}~\bibnamefont{Mukamel}},
  \bibinfo{journal}{Phys. Rev. B} \textbf{\bibinfo{volume}{69}},
  \bibinfo{pages}{155430} (\bibinfo{year}{2004}).

\bibitem[{\citenamefont{Sai et~al.}(2005)\citenamefont{Sai, Zwolak, Vignale,
  and Ventra}}]{sai}
\bibinfo{author}{\bibfnamefont{N.}~\bibnamefont{Sai}},
  \bibinfo{author}{\bibfnamefont{M.}~\bibnamefont{Zwolak}},
  \bibinfo{author}{\bibfnamefont{G.}~\bibnamefont{Vignale}}, \bibnamefont{and}
  \bibinfo{author}{\bibfnamefont{M.~D.} \bibnamefont{Ventra}},
  \bibinfo{journal}{Phys. Rev. Lett.} \textbf{\bibinfo{volume}{94}},
  \bibinfo{pages}{186810} (\bibinfo{year}{2005}).

\bibitem[{\citenamefont{Kosov}(2004)}]{kosov}
\bibinfo{author}{\bibfnamefont{D.~S.} \bibnamefont{Kosov}},
  \bibinfo{journal}{J. Chem. Phys.} \textbf{\bibinfo{volume}{120}},
  \bibinfo{pages}{7165} (\bibinfo{year}{2004}).

\bibitem[{\citenamefont{Burke et~al.}(2005)\citenamefont{Burke, Car, and
  Gebauer}}]{burke}
\bibinfo{author}{\bibfnamefont{K.}~\bibnamefont{Burke}},
  \bibinfo{author}{\bibfnamefont{R.}~\bibnamefont{Car}}, \bibnamefont{and}
  \bibinfo{author}{\bibfnamefont{R.}~\bibnamefont{Gebauer}},
  \bibinfo{journal}{Phys. Rev. Lett.} \textbf{\bibinfo{volume}{94}},
  \bibinfo{pages}{146803} (\bibinfo{year}{2005}).

\bibitem[{\citenamefont{Kohn}(1964)}]{kohn}
\bibinfo{author}{\bibfnamefont{W.}~\bibnamefont{Kohn}}, \bibinfo{journal}{Phys.
  Rev.} \textbf{\bibinfo{volume}{133}}, \bibinfo{pages}{A171}
  (\bibinfo{year}{1964}).

\bibitem[{\citenamefont{Nitzan}(2001)}]{nitzanreview}
\bibinfo{author}{\bibfnamefont{A.}~\bibnamefont{Nitzan}},
  \bibinfo{journal}{Ann. Rev. Phys. Chem.} \textbf{\bibinfo{volume}{52}},
  \bibinfo{pages}{681} (\bibinfo{year}{2001}).

\bibitem[{\citenamefont{Mukamel}(1995)}]{mukamel}
\bibinfo{author}{\bibfnamefont{S.}~\bibnamefont{Mukamel}},
  \emph{\bibinfo{title}{Principles of Nonlinear Optical Spectroscopy}}
  (\bibinfo{publisher}{Oxford Univ. Press}, \bibinfo{year}{1995}).

\bibitem[{\citenamefont{Boyd}(2003)}]{boyd}
\bibinfo{author}{\bibfnamefont{R.~W.} \bibnamefont{Boyd}},
  \emph{\bibinfo{title}{Nonlinear Optics}} (\bibinfo{publisher}{Academic
  Press}, \bibinfo{year}{2003}).

\bibitem[{\citenamefont{Buttiker}(1985)}]{buttiker}
\bibinfo{author}{\bibfnamefont{M.}~\bibnamefont{Buttiker}},
  \bibinfo{journal}{Phys. Rev. B} \textbf{\bibinfo{volume}{32}},
  \bibinfo{pages}{1846} (\bibinfo{year}{1985}).

\bibitem[{\citenamefont{Aviram and Ratner}(1974)}]{aviramratner}
\bibinfo{author}{\bibfnamefont{A.}~\bibnamefont{Aviram}} \bibnamefont{and}
  \bibinfo{author}{\bibfnamefont{M.~A.} \bibnamefont{Ratner}},
  \bibinfo{journal}{Chem. Phys. Lett.} \textbf{\bibinfo{volume}{29}},
  \bibinfo{pages}{274} (\bibinfo{year}{1974}).

\bibitem[{\citenamefont{Troisi and Ratner}(2002)}]{troisi}
\bibinfo{author}{\bibfnamefont{A.}~\bibnamefont{Troisi}} \bibnamefont{and}
  \bibinfo{author}{\bibfnamefont{M.~A.} \bibnamefont{Ratner}},
  \bibinfo{journal}{J. Am. Chem. Soc.} \textbf{\bibinfo{volume}{124}},
  \bibinfo{pages}{14528} (\bibinfo{year}{2002}).

\bibitem[{\citenamefont{Maldague}(1977)}]{maldague}
\bibinfo{author}{\bibfnamefont{P.~F.} \bibnamefont{Maldague}},
  \bibinfo{journal}{Phys. Rev. B} \textbf{\bibinfo{volume}{16}},
  \bibinfo{pages}{2437} (\bibinfo{year}{1977}).

\bibitem[{\citenamefont{Davis et~al.}(1997)\citenamefont{Davis, Wasielewski,
  Ratner, Mujica, and Nitzan}}]{davis}
\bibinfo{author}{\bibfnamefont{W.~B.} \bibnamefont{Davis}},
  \bibinfo{author}{\bibfnamefont{M.~R.} \bibnamefont{Wasielewski}},
  \bibinfo{author}{\bibfnamefont{M.~A.} \bibnamefont{Ratner}},
  \bibinfo{author}{\bibfnamefont{V.}~\bibnamefont{Mujica}}, \bibnamefont{and}
  \bibinfo{author}{\bibfnamefont{A.}~\bibnamefont{Nitzan}},
  \bibinfo{journal}{J. Phys. Chem.} \textbf{\bibinfo{volume}{101}},
  \bibinfo{pages}{6158} (\bibinfo{year}{1997}).

\bibitem[{\citenamefont{Liang et~al.}(2004)\citenamefont{Liang, Gosh, Paulsson,
  and Datta}}]{dattaprofiles}
\bibinfo{author}{\bibfnamefont{G.~C.} \bibnamefont{Liang}},
  \bibinfo{author}{\bibfnamefont{A.~W.} \bibnamefont{Gosh}},
  \bibinfo{author}{\bibfnamefont{M.}~\bibnamefont{Paulsson}}, \bibnamefont{and}
  \bibinfo{author}{\bibfnamefont{S.}~\bibnamefont{Datta}},
  \bibinfo{journal}{Phys. Rev. B} \textbf{\bibinfo{volume}{69}},
  \bibinfo{pages}{115302} (\bibinfo{year}{2004}).

\end{thebibliography}

\end{document}